\documentclass[a4paper]{article}
\usepackage{INTERSPEECH2014, epsfig}
\usepackage{amsmath, amssymb, bm, graphicx, url}
\usepackage{pdfpages}

\ninept
\def\reg{{\rm\ooalign{\hfil
     \raise.07ex\hbox{\scriptsize R}\hfil\crcr\mathhexbox20D}}}

\def\X{{\cal X}}
\def\Y{{\cal Y}}
\def\R{\mathbb{R}}

\def\MFCD{\mbox{MFC-$\Delta\,$}}
\def\MFSD{\mbox{MFS-$\Delta\,$}}
\def\MFBD{\mbox{MFB-$\Delta\,$}}   
\def\PLPD{\mbox{PLP-$\Delta\,$}}

\makeatletter

\DeclareRobustCommand\onedot{.}

\def\eg{\emph{e.g}\onedot} 
\def\ie{\emph{i.e}\onedot} 
 
\def\etc{\emph{etc}\onedot}

\makeatother

\newcommand{\capmar}{-0.3cm}

\newcommand{\secmar}{-0.1cm}
\newcommand{\subsecmar}{-0.15cm}

\title{Learning An Invariant Speech Representation}    

\makeatletter
\def\name#1{\gdef\@name{#1\\}}
\makeatother 

\name{{\em Georgios Evangelopoulos$^{\star \dagger}$, Stephen Voinea$^{\star}$, Chiyuan Zhang$^{\star}$, Lorenzo Rosasco$^{\star \dagger}$, Tomaso Poggio$^{\star \dagger}$\thanks{This material is based upon work supported by the Center for Brains, Minds and Machines (CBMM), funded by NSF STC award CCF-1231216. Lorenzo Rosasco acknowledges the financial support of the Italian Ministry of Education, University and Research FIRB project RBFR12M3AC.}}}
\address{$^{\star}$~Center for Brains, Minds and Machines $|$ McGovern Institute for Brain Research at MIT \\
  $^{\dagger}$~LCSL, Istituto Italiano di Tecnologia and Massachusetts Institute of Technology \\
{\small \tt [gevang, voinea, chiyuan, lrosasco]@mit.edu, tp@ai.mit.edu}}

\begin{document}

\includepdf[pages={1}]{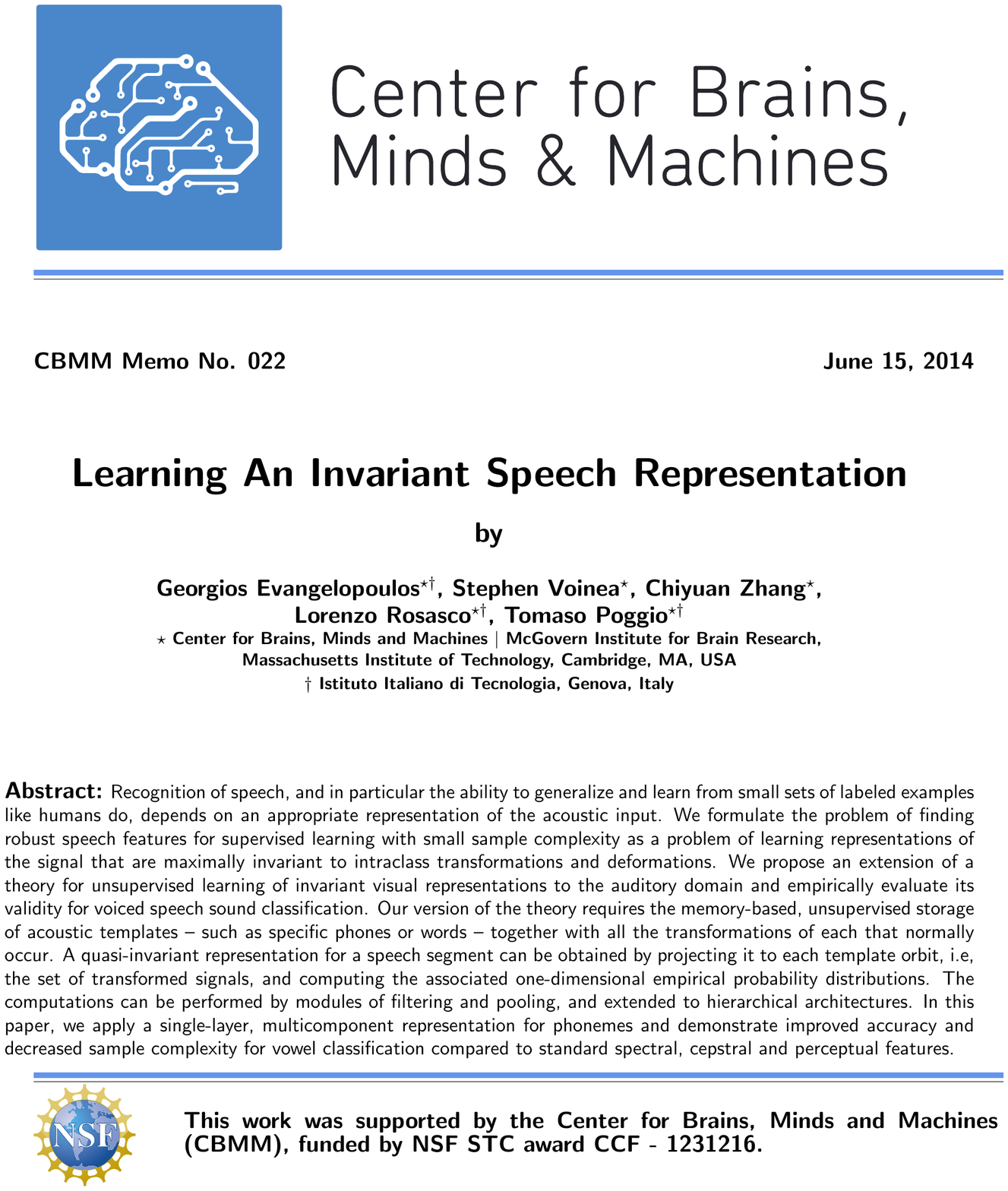}
\setcounter{page}{1}

\maketitle

\begin{abstract}


Recognition of speech, and in particular the ability to generalize and learn from small sets of labeled examples like humans do, depends on an appropriate representation of the acoustic input. We formulate the problem of finding robust speech features for supervised learning with small sample complexity as a problem of learning representations of the signal that are maximally invariant to intraclass transformations and deformations. We propose an extension of a theory for unsupervised learning of invariant visual representations to the auditory domain and empirically evaluate its validity for voiced speech sound classification. Our version of the theory requires the memory-based, unsupervised storage of acoustic templates -- such as specific phones or words -- together with all the transformations of each that normally occur. A quasi-invariant representation for a speech segment can be obtained by projecting it to each template orbit, \ie{}, the set of transformed signals, and computing the associated one-dimensional empirical probability distributions. The computations can be performed by modules of filtering and pooling, and extended to hierarchical architectures. In this paper, we apply a single-layer, multicomponent representation for phonemes and demonstrate improved accuracy and decreased sample complexity for vowel classification compared to standard spectral, cepstral and perceptual features.



\end{abstract}
\noindent{\bf Index Terms}: speech representation, invariance, acoustic features, representation learning, speech classification 


\section{Introduction}
\label{sec:intro}

The recognition of speech sounds and units (phonemes, words) from the acoustic input, and the resilience of humans to the range and diversity of speech variations, such as rate, mode and style of speech, speaker state, identity, or pronunciation \cite{Benzeghiba2007, Meyer2011}, might be related to the representation in the auditory cortex and ventral auditory pathway \cite{DeWitt2012}. Given a representation that is \emph{invariant} to changes in the signal that do not affect the perceived class, the task of recognition can be reduced to learning from a few examples \cite{Anselmi2014}. Hierarchical, cortex-inspired models for selective and invariant visual representations, have shown to be predictive of human performance and competitive to computer vision systems  \cite{Serre2007, Boureau2010}. 
Their main principle, of re-using modules of fitering and pooling in multilayered hierarchies, 
is also the essence behind deep convolutional learning networks, recently applied for speech and sound recognition \cite{lee2009unsupervised, Anden2013, Abdel-Hamid2013, Hinton2012}. The empirical successes motivate different theoretical formalisms on what constitutes an effective data representation and how can one be learned in brains and machines \cite{Smale2009, Mallat2012, Anselmi2014}. 



In the context of learning from data, an effective representation is related to the sample complexity of a recognition problem, \ie{}, the number of training examples required for a given generalization error \cite{Poggio2003}. 
In that respect, an appropriate representation should be invariant to intraclass transformations, such as affine time-frequency maps or pitch changes, discriminative for different classes of speech `objects', and stable to signal deformations (nonlinear warping). Invariance is a key concept towards machines that learn from few labeled examples. 
   

  
In this paper we propose a family of invariant representations for speech, based on a theory for unsupervised learning of invariances \cite{Liao2013}, where a segment is represented by the empirical distributions of the projections on sets of transformed template signals. The representation is quasi-invariant and discriminative for $C$ classes, when using a finite number of templates $K$ that depends on $C$, 
and reduces the sample complexity of a supervised classification stage \cite{Anselmi2014}. Assuming access to templates and their transformations, through unsupervised observation, and storage in the form of ``neuron" synapses or convolutional network weights, invariance can be learned in an unsupervised way. The implied computations (dot-products and cumulative nonlinearities) are biologically plausible, consistent with the pooling of complex over simple cells and may suggest a computational goal for modules of the auditory cortex.

\vspace{\secmar}
\section{Related Work}
\label{sec:back}
Features for speech recognition are conventionally based on frame-level spectral representations, passed through one or multiple layers transformations including: frequency warping (mel or bark), compression (log, power-low), predictive coding, cosine transforms (cepstrum), normalization (\eg{}, loudness) \cite{Schatz2013, O'Shaughnessy2013}. Paths in this processing chain account for the mel-frequency cepstral (MFC) and perceptual linear predictive (PLP) coefficients. 
%
%
%
Robustness has been sought through representations inspired by properties of the early and central auditory cortical processing \cite{Stern2012}. Examples include perceptual critical bands \cite{Allen2009}, long-term temporal analysis through the modulation spectrum \cite{Greenberg1997} or temporal envelopes of spectral bands \cite{HermanskySharma1998}, the auditory spectrogram \cite{Chi2005} and spectrotemporal receptive fields \cite{Mesgarani2006} for multiresolution (frequency/rate/scale) analysis, 
localized time-frequency modulation tuning \cite{Kleinschmidt2003, Bouvrie2008}, and hierarchical organization \cite{Heckmann2011}. The representational effect of the transformed feature maps is to ignore irrelevant and noisy components of the speech signal, impose invariance by smoothing out nonphonetic variations and integrating information across time/frequency selective regions.      

Speech recognition witnessed performance gains by replacing feature extraction and acoustic modeling with 
multilayered, deep networks \cite{Abdel-Hamid2012, Hinton2012}. Convolutional networks \cite{Abdel-Hamid2013, Sainath2013}, which 
interchange convolution and pooling layers for local time \cite{lee2009unsupervised} or frequency \cite{Abdel-Hamid2012} invariance, result in distributed representations that are tolerant to speech variability (speaker and noise) in a data driven way \cite{Yu2013a}. 
%
Their performance though depends on abundant labeled examples, for learning the filters and dependencies in a supervised way. 
A scattering transform representation \cite{Mallat2012} builds shift-invariance and warp-stability over long temporal windows 
through cascades of analytic wavelet convolutions and modulus operators \cite{Anden2013}. 
Our proposed invariant representation is learned in a feed-forward, unsupervised mode, using stored filter-templates, does not explicitly model the signal and is \emph{feature-agnostic}, \ie{} it can be derived from any base speech representation.     


\vspace{\secmar}

\section{Group-Invariant Speech Representations}

{\bf Transformations, Invariance and Learning:} Let $\Phi(s)$ be a representation of a one-dimensional, 
time-varying signal $s \in \X $, with $\X\!\subset\!\R^d$, or $L^2(\R)$, the $d$-dimensional representation space indexed by time, frequency or (any) coding coefficient samples. For a signal processed or mapped from the raw sound waveform to some feature space $\Phi_0(s)$, $\X$ can be, for example the space of spectral coefficients or spectro-temporal patches. The dimension $d$ depends on the length of the represented speech unit (frame, phone, sub-word, word).  


Signal instances $s$ belong to some ground-truth class \mbox{$y\in \Y=\{1,\dots,C\}$}, depending on the task (\eg{}, phonetic, lexical, speaker class). Given a (training) set of $n$ pairs $\{(s_i,y_i)\}_{i=1}^{n} \in (\X, \Y)$, sampled from an unknown distribution, the supervised learning problem, posed here as a multiclass classification task, is to estimate a function $f\!\!:\!\X \rightarrow \Y$ which predicts or assigns label $y \in \Y$ in a sample $s \in \X$. 
A representation or map $\Phi(s)\!\!:\!\X \rightarrow \R^d$, is said to be be discriminative (or unique with respect to individual class members), when \mbox{$\Phi(s) \neq \Phi(s')\!\Leftrightarrow\!f(s) \neq f(s')$}, invariant to intraclass variations caused by class-preserving transformations $s'=gs$ when \mbox{$\Phi(s) = \Phi(s')\!\Leftrightarrow\!f(s') = f(s)$}, and stable, \ie{}, Lipschitz continuous when \mbox{$d(\Phi(s), \Phi(s')) \leq L \|s-s'\|_2$}, for some metric $d(\cdot)$ and $L\leq1$. An invariant representation $\Phi(s)$ on $\X$ reduces the number $n$ of labeled examples required for learning $f$ with a given generalization error \cite{Anselmi2014, Poggio2003}. 

A large family of typical speech transformations and deformations can be described by $s'(t) = v(t)*s(\phi(t)) + n(t)$, where $*$ denotes convolution/filtering by $v(t)$, $n(t)$ is noise, and $\phi(t)$ a linear or nonlinear function that applies mapping $t \mapsto \phi(t)$ on signal $s(t)$. These include affine maps (translation and scaling $\phi_{a,\tau}(t) = at - \tau$, nonlinear time-warping, and more complex, domain-dependent changes like pronunciation, dialect, context, emotion etc. They can occur or mapped in time $s(t)$, frequency $S(\omega)$ or joint spectro-temporal domains.  



\vspace{\subsecmar}
\subsection{Modules For Invariant Representations}
\label{sec:nod}
Transformations of a signal $s$ can be described through a group $G$ (\eg{} the affine group in $\R$), as the result of the action of an element $g \in G$ on the signal $gs(t) = s(g^{-1}t)$ \cite{Mallat2012}. A representation $\Phi(s)$ is said to be invariant to group $G$, if the action of any $g \in G$ does not have an effect on it:  
\begin{equation}
\Phi(gs) = \Phi(s), \forall g \in G   
\end{equation}
The set of transformed signals $g s, \forall g\in G$, generated under the action of $G$ on $s$, is the {\emph group orbit}
\begin{equation} 
O_s=\{gs\in\R^d |\, g\in G \},
\end{equation}
with an \emph{equivalence relation} defined by $G$ for signals that belong to the same orbit: $s \sim s'\!\Leftrightarrow\!\exists g \in G\!\!: s' = gs$. 
The orbit is then an invariant representation of $s$, since $O_s = O_{gs}, \forall g \in G$. An empirical measure of $O_s$ is the high-dimensional probability distribution $P_s, \forall gs \in O_s$, which is \emph{invariant} and \emph{unique}. It can thus be used for comparing orbits, and by equivalence, the mapped signals: $s \sim s'\!\Leftrightarrow\!O_s \sim O_s'\!\Leftrightarrow\!P_s \sim P_s'$. 


A metric on $d$-dimensional distributions $P_s$ can be approximated, following Cram{\'e}r-Wold Theorem and concentration of measures \cite{Anselmi2014}, through a finite number of one-dimensional distributions $\{P_{\langle s, t^k \rangle}\}_{k=1}^{K}$ induced from the projections of $gs$ on $K$ \emph{template signals} $\{t^k\}_{k=1}^{K}$, where $\langle \cdot, \cdot \rangle$ is a dot-product defined in signal space $\X$. Estimation of the distribution of projections on a single $t^k$ requires access to all signal transformations $gs$ under $G$. For unitary groups, and normalized dot-products, \ie{} 
$\langle s, t^k \rangle/\left(\|s\|\|t^k\|\right)$, the property   
\begin{equation}
\langle gs, t^k \rangle = \langle s, g^{-1} t^{k} \rangle,
\end{equation}
allows for obtaining the one-dimensional distribution for $t^k$ through all projections of $s$ on the transformations of $t^k$ under $G$. The main implication is that an invariant representation can be learned in a memory based, unsupervised way from stored, transformed versions of a large but finite number of templates. 


An estimate of the empirical distribution for a finite group $G$ is obtained by the cumulative summation through a set of $N$ nonlinear functions $\{\eta_n(\cdot)\}_{n=1}^{N}$: 
\begin{equation}
\mu_n^k(s) = \frac{1}{|G|}\sum_{j=1}^{|G|} \eta_n \left( \langle s,g_j^{-1} t^k \rangle \right) 
\label{eq:dist}
\end{equation}
Depending on the form of $\eta(\cdot)$, the above estimate is the cumulative distribution function (\ie{}, smooth sigmoid functions for $N$ bins) or any moment (mean, second-order energy, approx. max \etc). Note that, the set of all moments determines a distribution of bounded-support as in $P_{\langle s, t^k \rangle}$. Equation (3) describes a generic pooling mechanism over dot-products with functions $gt^k$, that is invariant to the actions of the group as a group average. The final signature is the collection of all estimates, for all $K$ templates, $\Phi(s) = (\{\mu_n^1(s)\},\ldots,\{\mu_n^k(s)\}) \in \R^{NK}$ or: 
\begin{equation}
\Phi(s) = (\mu_1^1(s),\ldots,\mu_N^1(s),\ldots, \mu_1^k(s),\ldots,\mu_N^k(s)). 
\label{eq:rep}
\end{equation}
By extension, \emph{partial invariance} to any complex, but smooth transformation $G$, can be achieved for sufficiently localized dot-products (partially-observable groups), through the local linearization of the transformation \cite{Anselmi2014}.  
    
\vspace{\subsecmar} 
\subsection{Obtaining Templates and Orbit Samples}

The template signals $t^k$ can in principle be random and even unrelated to the input speech segment (for the case of affine transformations), \eg{}, coming from different phonetic categories. The number of templates required for discriminating $C$ classes is $K\geq (2/c\epsilon^{2})\log(C/\delta)$, for an $\epsilon$-approximation error of the true distributions by the $K$ empirical estimates with confidence $1-\delta^2$. 
The theory requires stored templates and their transformations, which involves no supervision 
and is equivalent to setting the weights of the representation architecture.         
The transformations for a given template $t^k$, or equivalently, the samples from the transformation orbit $O_t^k$, can be acquired, through unsupervised observation via the assumption of temporal adjacency: sounds close together in time correspond with high probability to the same source undergoing a transformation (\ie{}, speech changing in time, the phonetic variations of a speaker, or the examples provided to children by caregivers during development and language learning \cite{Roy2009a}). This is the biological equivalent of sampling the class-specific subspace of $\X$ and storing samples from the orbits $O_t^k$ of template $t^k$ in an unsupervised way, \ie{}, the label of $t^k$ is not important. 
 
%


For learning speech representations for machine-based recognition, the templates and transformations can also be stored (like in a pre-trained network), or obtained by transforming a set of randomly selected templates \cite{Liao2013}. This belongs to a family of techniques for augmenting training datasets for machine learning \cite{Niyogi1998, Jaitly2013}.    
%
%
Given unlabeled but structured data (\ie{}, belonging to different sound categories, speaker or dialect classes), clustering or spectral clustering can be used for forming partitions that correspond to smoothly varying signals, approximating samples from smooth transformation groups. Given a few examples (templates) $t^k$, online learning, and bootstrapping estimates of the empirical distributions, can be used for progressively adding examples to the orbit sets $\{g_jt^k\}$. Observations that are not added can form new orbit sets. Templates can be chosen from a set different from the learning set, using \emph{categorical grouping}, for example picking samples from different phones, speakers, dialects, etc. In this paper (Sec.~\ref{sec:exp_temp}) we explore both using phonetic category groups (templates are different vowels) and clustering.    






\vspace{\secmar}
\section{Computing Segment Representations}
\label{sec:comp}

Re-using pooling-filtering modules with templates of smaller support and adding layers on top of representation (\ref{eq:rep}), can increase the range of invariance 
and selectivity to larger signal parts \cite{Anselmi2014, Smale2009}. For this paper, we consider a single-layer representation at the level of segmented phones (50-200ms), that has multiple components (one per template). It assumes that templates and input are in the same space and have the same support. A fixed-length base representation, that handles varying lengths, is obtained by local averaging of features across frames (Sec. \ref{sec:exp_feats}). An interpretation, in terms of our model, is that waveforms are mapped on a base (0-th or 1-st) feature layer through (adaptive) time and frequency pooling (filterbanks). 



Consider an input set $X = \{x_i\}_{i=1}^{n} \in \X\!\subset\!\R^d$, where $x_i = \Phi_0(s_i)$ the $d$-dimensional base representation of the segment. For each $x \in X, t \in \X$, the dot-product $\langle x, t^k \rangle$ is given by $x^T t$, as $\X\!\subset\!\R^d$. 
Assume access to a different set $T \in \X$, denoted as the \emph{template set}, to be used as a pool of data (observations) for obtaining templates $t_k$ and samples from their orbit sets, \ie{}, transformations $t_{j}^k=\{g_{j}t^{k}\}$ for arbitrary and unknown $\{g_{j}\}$. The template dataset $T$ is partitioned into $K$ non-overlapping subsets $T_k = \{t_{j}^{k}\}_{j=1}^{|T_k|}, T = \bigcup_{k=1}^{K} T_k$, using categorical grouping or distance-based/spectral clustering. Under such a meaningful partition, that preserves some structure in the data, we approximate Eq.~(\ref{eq:dist}) by $\mu_n^k(x) \approx |T_k|^{-1}\sum_{j=1}^{|T_k|} \eta_n \left( \langle x, t_j^k \rangle \right)$. 


For each input instance $x_i$, the representation is obtained in three steps: 1)~{\bf Filtering}: estimate the normalized dot-product between $x_i$ and template $t_{j}^k$, for each template in $T_k$ and each of the $K$ sets, 2)~{\bf Pooling}: apply nonlinear functions (histograms) on the projection values within each $T_k$, 3)~{\bf Signature:} concatenate components from all subsets. If matrix ${\mathbf X} = (x_1,\ldots , x_n)^T$ holds the row-wise arrangement of samples, ${\mathbf T_k}=(t_{1}^k,\ldots, t_{|T_k|}^{k})^T$, $\tilde{\mathbf{X}}, \tilde{\mathbf{T}}$ the zero-mean/unit-norm normalized matrices, row- and column-wise respectively, and $\mathbb{I}_k$ the $|T| \times 1$ indicator vector for subset $k$, the computations can be written: 
$\tilde{\mathbf{X}} \tilde{\mathbf{T}}_k^T$ (projection on the set of one template $k$), $\tilde{\mathbf{X}}\tilde{\mathbf{T}}^T$ (projection on all $K$ template sets), $\eta^k(\tilde{\mathbf{X}}[\tilde{\mathbf{T}}_1^T, \ldots,\tilde{\mathbf{T}}_K^T]\mathbb{I}_k)$ (pooling over the elements of a set $k$). 

\begin{figure}
\centerline{
\includegraphics[width=0.64\columnwidth]{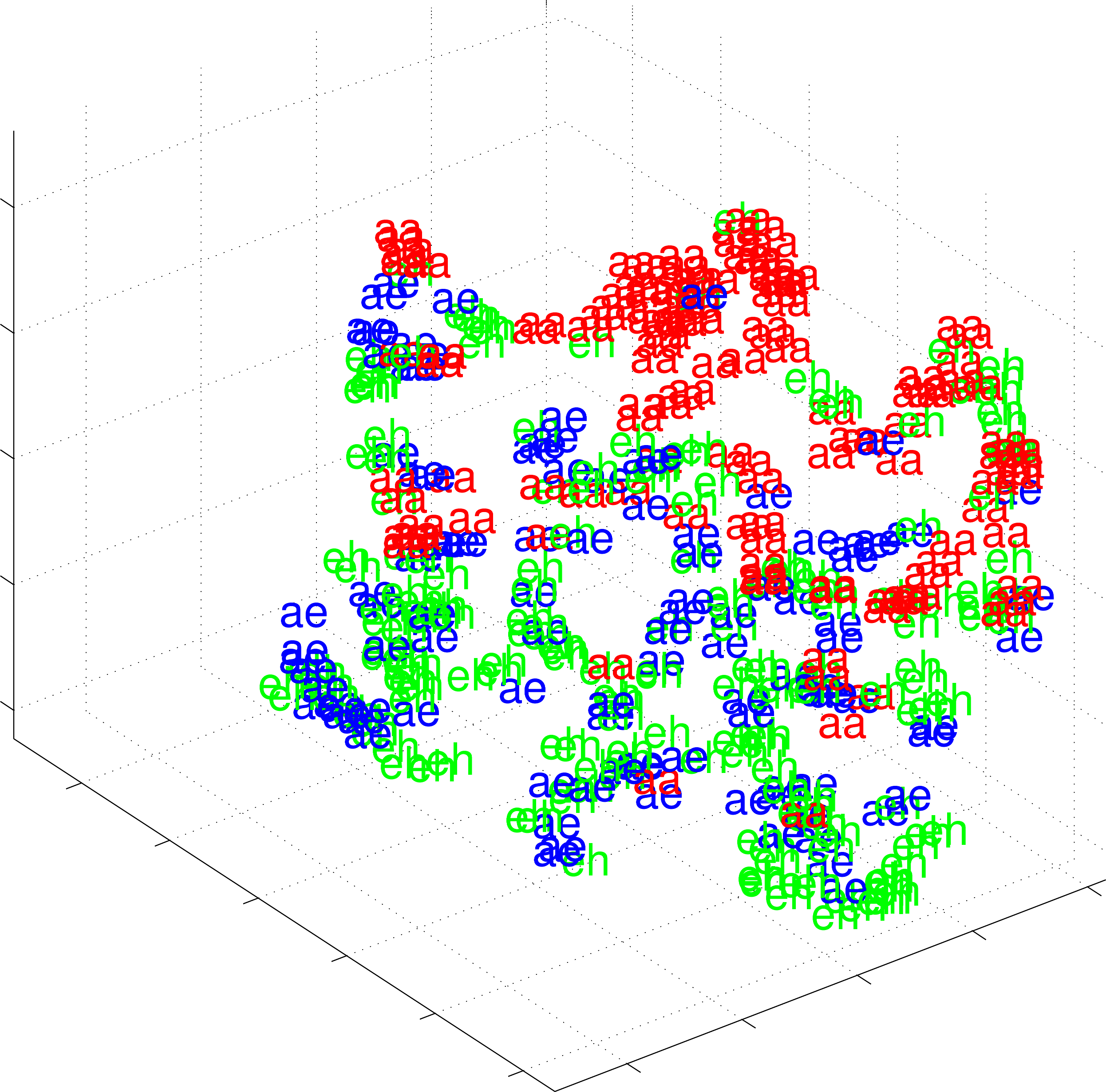}
\includegraphics[width=0.36\columnwidth]{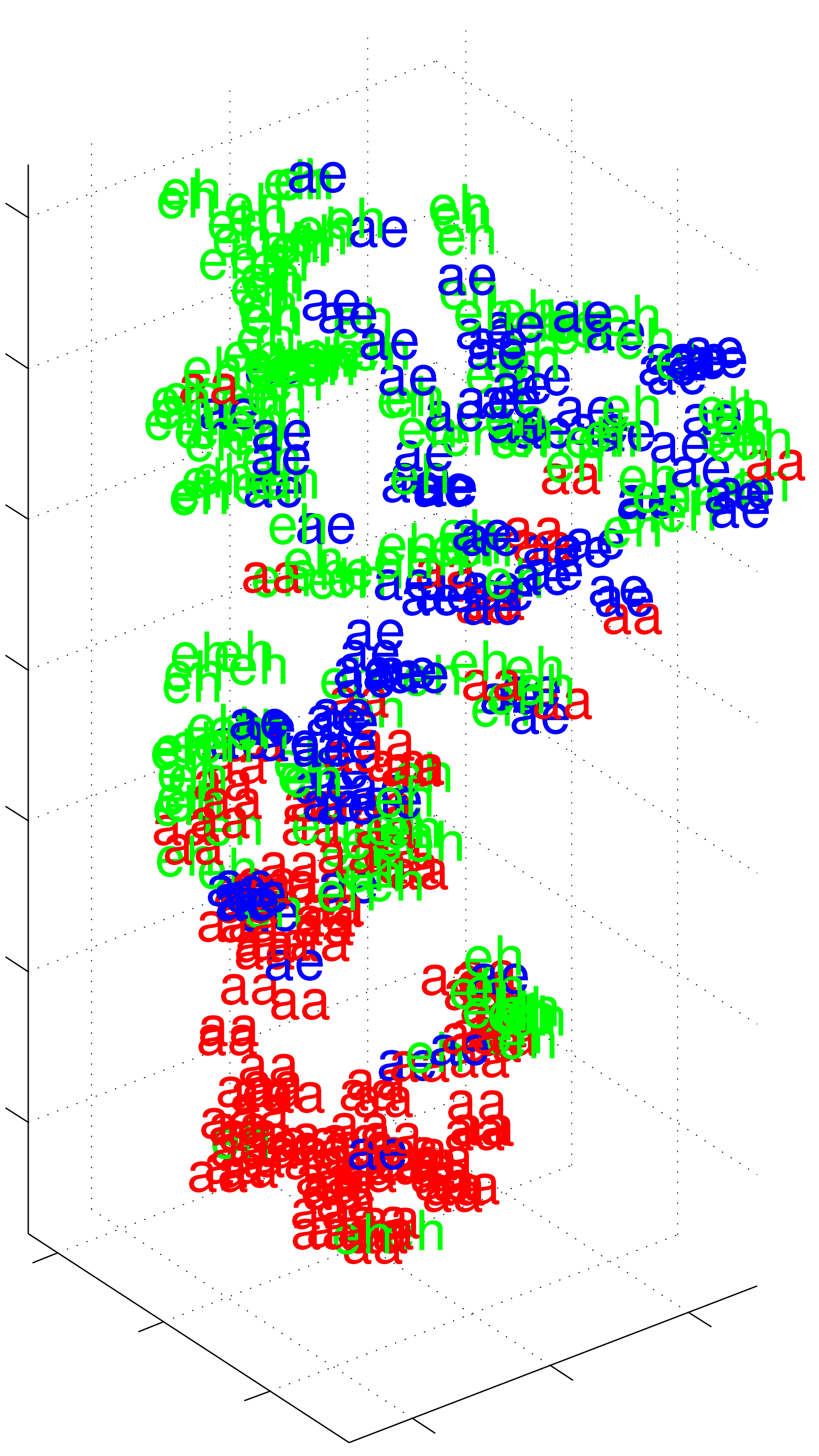}
}
\vspace{\capmar}
\caption{Vowels (/ae/, /aa/, /eh/) of TIMIT Core Set in base \PLPD (left) and proposed invariant representation space (right), shown as three-dimensional t-SNE embeddings \cite{Maaten2008}.}
\end{figure}

\vspace{\secmar}
\section{Vowel Classification}

\begin{table*}[!t!]
\begin{center}
\renewcommand*{\arraystretch}{1.1}
\eightpt  
\setlength{\tabcolsep}{2pt}
\begin{tabular}{c|l c r c r c c || l c r c r c c}
RLS& & $|T|$& $K$ & Groups & Dim  &  bER ($\%$) & ER ($\%$)   &  &$|T|$ & $K$ & Groups & Dim &  bER ($\%$)   & ER ($\%$)  \\\hline                  
OVA& \MFSD$\!$      &     &    &            & 615 & $ 57.72 $ & $ 48.95 $ & \MFBD$\!$       &     &    &            & 615 & $ 52.72 $ & $ 45.15 $\\ 
OVA& InvR(\MFSD$\!$)& Dev & 20 &{\bf phn}   & 400 & $ 56.02 $ & $ 49.47 $ & InvR(\MFBD$\!$) & Dev & 20 &{\bf phn}   & 400 & $ 53.53 $ & $ 47.71 $\\  
OVA& InvR(\MFSD$\!$)& Res & 20 &{\bf phn}   & 400 & $ 54.98 $ & $ 48.91 $ & InvR(\MFBD$\!$) & Res & 20 &{\bf phn}   & 400 & $ 48.77 $ & $ 44.47 $\\ 
OVA& InvR(\MFSD$\!$)& Res &135 &{\bf phn-dr}&2700 & $ \bm{51.67} $ & $ \bm{45.36} $ & InvR(\MFBD$\!$) & Res &135 &{\bf phn-dr}&2700 & $ \bm{47.85} $ & $ \bm{43.06} $\\ 
OVA& InvR(\MFSD$\!$)& Res &120 &{\bf kmeans}&2400 & $ 55.84 $ & $ 47.24 $ & InvR(\MFBD$\!$) & Res &120 &{\bf kmeans}&2400 & $ 55.10 $ & $ 46.60 $\\
\hline \hline
OVA& \MFCD$\!$      &     &    &            & 195 & $ 54.08 $ & $ 45.92 $ & \PLPD$\!$       &     &    &            &195 & $ 53.52 $ & $ 45.24 $\\ 
OVA& InvR(\MFCD$\!$)& Dev & 20 &{\bf phn}   & 400 & $ 45.67 $ & $ 41.82 $ & InvR(\PLPD$\!$) & Dev & 20 &{\bf phn}   &400 & $ 45.66 $ & $ 41.99 $\\ 
OVA& InvR(\MFCD$\!$)& Res & 20 &{\bf phn}   & 400 & $ 41.56 $ & $ 38.66 $ & InvR(\PLPD$\!$) & Res & 20 &{\bf phn}   &400 & $ 43.04 $ & $ 39.64 $\\  
OVA& InvR(\MFCD$\!$)& Res &135 &{\bf phn-dr}&2700 & $ \bm{41.10} $ & $ \bm{37.80}$ & InvR(\PLPD$\!$) & Res &135 &{\bf phn-dr}&2700& $ \bm{42.30} $ & $\bm{38.53}$\\
OVA& InvR(\MFSD$\!$)& Res &200 &{\bf kmeans}&4000 & $ 46.48 $ & $ 41.61 $ & InvR(\PLPD$\!$) & Res &200 &{\bf kmeans}&4000& $ 45.75 $ & $ 41.31 $\\
\hline 
\end{tabular}
\end{center}
\vspace{\capmar}
\caption{\small Vowel classification error (ER) and balanced error rate (bER) using different base representations $\Phi_0(s)$, the proposed \mbox{$\Phi(s)$ = InvR($\Phi_0$)} and linear Regularized Least Squares  classifiers (OVA: One-vs-All).} 
\label{tab:res}
\end{table*}

To demonstrate the potential of the proposed speech representation, 
we present comparisons with standard spectral and cepstral features for an acoustic classification task on TIMIT dataset. We focus on vowels, including dipthongs, the largest and most challenging phonetic subcategory on the set \cite{Halberstadt1997}. For the task of classifying pre-segmented phone instances in 20 vowel classes, we use supervised linear classifiers without post-mapping the vowel categories prior to scoring. 
We train multiclass Regularized Least Squares (RLS) with linear kernels \cite{Tacchetti2013}, using $1/6$ of the training set for validating the regularization parameter. The use of a kernel-based, nonlinear RLS is expected to further improve the classification accuracy \cite{Karsmakers2007,Rifkin2007}. 

%

We used the standard Train (462 speakers, 45572 vowel tokens) and Test (168 speakers, 16580 vowel tokens) partitions and evaluate performance through phone error rates on the Core set (24 speakers, 2341 vowel tokens). For templates and orbit sets, we use either the standard development subset (50 speakers, 4938 vowel tokens), which we denote as ``Dev" or the remaining Test excluding Core (118 speakers, 14239 vowel tokens), which we denote as ``Res". 
Note that, the distribution of tokens across the 20 vowel classes is similar on all sets.   




\vspace{\subsecmar}
\subsection{Baseline Features and Base Representations}
\label{sec:exp_feats}

We used four base representations, extracted using HTK \cite{HTK09} at frame level from the segmented phones expanded by 30ms. An analysis window of 25ms at $16\rm kHz$, with 10ms shifts was used for frame-level encoding. Features were enhanced by first and second-order derivatives (\mbox{$\Delta\!-\!\Delta\Delta$}), to incorporate temporal context. Frames were aggregated across the phoneme duration by averaging on three segments (3-4-3 ratio) \cite{Rifkin2007, Chang2009} and on the two boundary extensions, resulting on a 5-component representation 
of dimension $d' = 5\times 3 \times d$, with $d = |\Phi_0(\cdot)|$ the length of the frame level features.
%
%
%
%
Four types of coefficients where used for encoding: mel-frequency spectral (MFS, $d=41$), log-MFS filterbank (MFB, $d=41$), mel-frequency cepstral (MFC, $d=13$) and perceptual linear prediction (PLP, $d=13$), with standard filterbank and cepstral parameters: 40 mel-bank for MFS (plus frame energy), 26 mel-bank for MFC/PLP (order 12), keeping 13 coefficients (including 0th). 

    
\vspace{\subsecmar}
\subsection{Template Sets and Signature Derivation}
\label{sec:exp_temp}

As a case study, we consider templates from the same phonetic categories (/vowels/) and samples of ``transformed templates'' obtained from a \emph{template set} not used for learning, in our case either the ``Dev" or the ``Res" TIMIT sets (the latter roughly three times larger). All instances in the set are assigned to a group, either based on categorical metadata (\eg{}, phone, word, phrase, dialect, speaker) or by clustering on the base feature space. Even though a form of labeling is used in the former, this serves only to simulate the unsupervised observation of the orbit samples (\eg{}, ``/aa/ instances of one dialect region"). We used $N$-bin histograms for the empirical distributions $\{\mu_n^k\}_{n=1}^{N}$ which outperformed in this task other moment sets. All reported results refer to $N\!=\!20$ bins. The final $KN$-dimensional signature $\Phi(s)$ in (\ref{eq:rep}) formed by concatenating $K$ histograms, each derived from the projections to all elements of a single set. All projections, as in Sec.~\ref{sec:comp}, are normalized 
dot-products and feature matrices are standardized (using the training set) before both the projections and the classification. 

\vspace{\subsecmar}
\subsection{Evaluations and Discussion}

Table~\ref{tab:res} shows classification results for the 20-class task on Core Set, listing both average and balanced error rates. Each block corresponds to a different base representation $\Phi_0(s)$ and the proposed invariant representation built on top of it, \ie{}, using dot-products on this space, is denoted by ${\rm InR}(\cdot)$. Results with three potential partitions of the template set $T$ on subsets are shown: phoneme category ({\bf phn}), phoneme and dialect category ({\bf phn-dr}) and {\bf kmeans}. The number of groups depend on $T$ for the first two ($K=20$ and $135$ resp.), while for the latter is a free parameter (here $K=120$ and $K=200$). 

\begin{figure}[t]
\centerline{
\includegraphics[width=0.98\columnwidth]{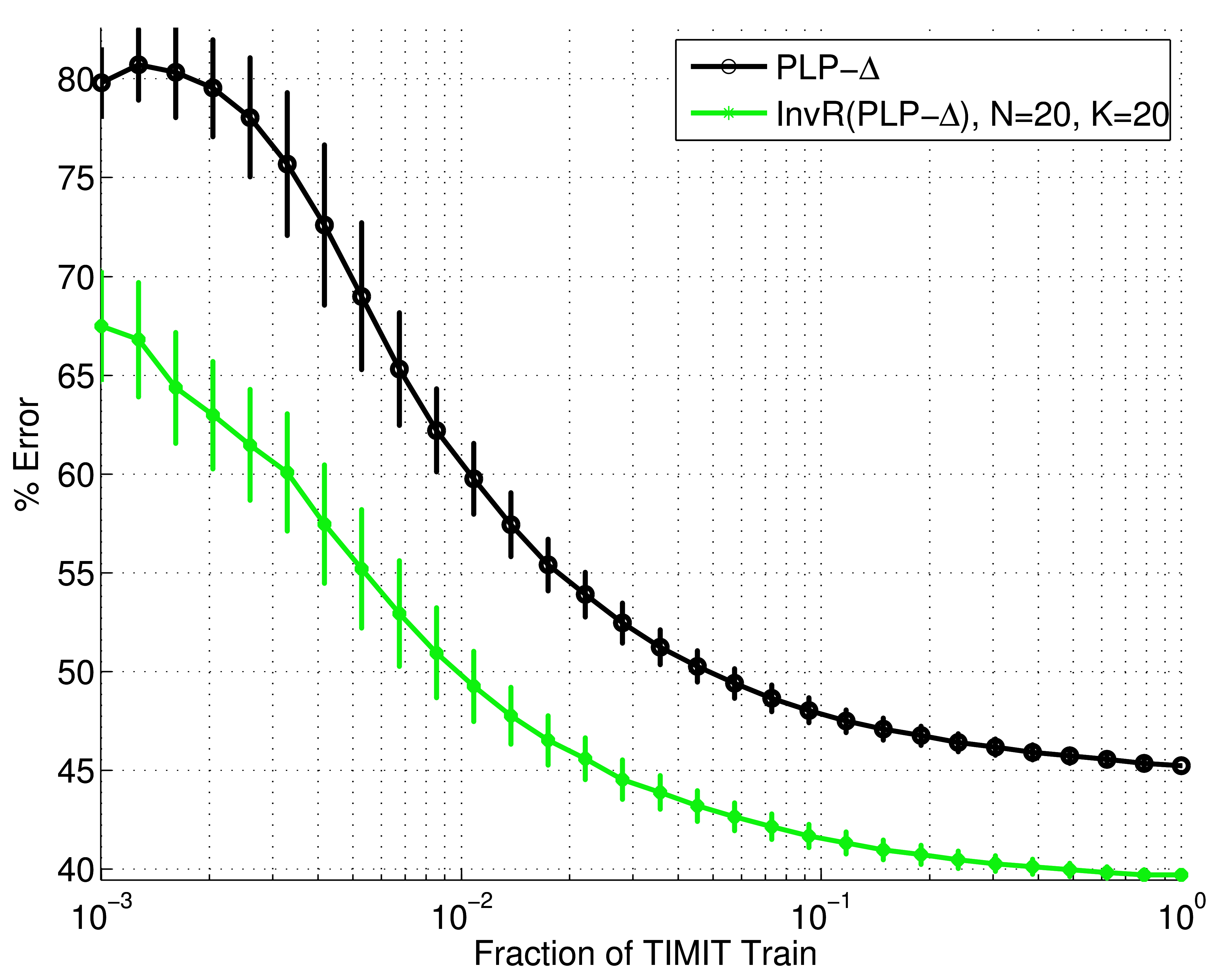}}
\vspace{\capmar}
\caption{\small Classification error dependency on the size of training set, decreasing TIMIT Train Set from $45572$ to $46$ (20-class, TIMIT Core Test). The plots correspond to the best baseline with \PLPD, with and without the invariant representation, averaged over 500 random partitions at each fraction level.}
\label{fig:drop}
\end{figure}

The invariant representation, using the richer ``Res" set, systematically improves accuracy with One-Vs-All (OVA) classifiers, with a decrease in error rates of 
$8.12\%$ (\MFCD) and $6.7\%$ (\PLPD). Using the smaller ``Dev" set improves accuracy in the case of the decorrelated features by $4.1\%$ for \MFCD and $3.25\%$ for \PLPD. Thus, the number of orbit samples (or transformed templates) is related to more accurate group averages, which is aligned with the theoretic requirements access to the entire orbit. For forming the orbit sets, the best results are obtained by using the phn-dr category, that assumes a template for one phonetic category over one dialect, giving 135 template orbit sets for ``Res". In this case, the set labels the 20 categories of the supervised task; the scheme provides the overall best accuracy ($37.80\%$ error for \MFCD). This performance is comparable to previously reported segmental-based classification results \cite{Bouvrie2008}, without using additional cues (vowel log-duration \cite{Karsmakers2007, Rifkin2007, Chang2009, Meyer2011}) and One-Vs-One (OVO) schemes. Results with the cosine-distance kmeans  clustering of the template set are better than baselines but lacking compared to the phn-dr, that seemingly  provides better orbit approximations. On sample complexity, the proposed representation attains performance comparable to the $37.55\%$ \PLPD performance on OVO (190 binary classifiers). Fig.~\ref{fig:drop} shows how the proposed representation is consistently better than standard PLP while decreasing the training set size down to three orders.    






\vspace{\secmar}
\section{Conclusions}
A novel representation for acoustic signals was proposed on the basis of a theory for learning representations, with formal assertions for invariance to simple and complex, but localized, transformations. The theory explains the properties of hierarchical architectures, specifies how to build representations of improved supervised learning capacity and conjectures a mode for their unsupervised learning in the auditory cortex. We empirically demonstrated that a single-layer, phone-level representation, inspired by these principles and extracted from base speech features, improves segment classification accuracy and decreases the number of training examples.  
We plan to evaluate multilayer, hierarchical extensions for larger-scale speech representations (\eg{}, words) on zero-resource matching \cite{Schatz2013} and the applicability for acoustic modeling in ASR. We are also working towards group-structure-aware unsupervised and online learning of the template orbit sets.     

\newpage 

\eightpt
\label{sec:refs}
\bibliographystyle{IEEEtran}
\bibliography{CBMM_Memo_022}

\begin{thebibliography}{10}
\providecommand{\url}[1]{#1}
\csname url@samestyle\endcsname
\providecommand{\newblock}{\relax}
\providecommand{\bibinfo}[2]{#2}
\providecommand{\BIBentrySTDinterwordspacing}{\spaceskip=0pt\relax}
\providecommand{\BIBentryALTinterwordstretchfactor}{4}
\providecommand{\BIBentryALTinterwordspacing}{\spaceskip=\fontdimen2\font plus
\BIBentryALTinterwordstretchfactor\fontdimen3\font minus
  \fontdimen4\font\relax}
\providecommand{\BIBforeignlanguage}[2]{{%
\expandafter\ifx\csname l@#1\endcsname\relax
\typeout{** WARNING: IEEEtran.bst: No hyphenation pattern has been}%
\typeout{** loaded for the language `#1'. Using the pattern for}%
\typeout{** the default language instead.}%
\else
\language=\csname l@#1\endcsname
\fi
#2}}
\providecommand{\BIBdecl}{\relax}
\BIBdecl

\bibitem{Benzeghiba2007}
M.~Benzeghiba, R.~{De Mori}, O.~Deroo, S.~Dupont, T.~Erbes, D.~Jouvet,
  L.~Fissore, P.~Laface, A.~Mertins, C.~Ris, R.~Rose, V.~Tyagi, and
  C.~Wellekens, ``Automatic speech recognition and speech variability: A
  review,'' \emph{Speech Communication}, vol.~49, no. 10-11, pp. 763--786, Oct.
  2007.

\bibitem{Meyer2011}
B.~T. Meyer, T.~Brand, and B.~Kollmeier, ``{Effect of speech-intrinsic
  variations on human and automatic recognition of spoken phonemes},''
  \emph{The Journal of the Acoustical Society of America}, vol. 129, no.~1, pp.
  388--403, Jan. 2011.

\bibitem{DeWitt2012}
I.~DeWitt and J.~P. Rauschecker, ``Phoneme and word recognition in the auditory
  ventral stream,'' \emph{Proc. of the National Academy of Sciences (PNAS)},
  vol. 109, no.~8, pp. E505--14, Mar. 2012.

\bibitem{Anselmi2014}
\BIBentryALTinterwordspacing
F.~Anselmi, J.~Z. Leibo, L.~Rosasco, J.~Mutch, A.~Tacchetti, and T.~Poggio,
  ``Unsupervised learning of invariant representations in hierarchical
  architectures,'' \emph{ArXiv}, Jan. 2014. [Online]. Available:
  \url{http://arxiv.org/abs/1311.4158}
\BIBentrySTDinterwordspacing

\bibitem{Serre2007}
T.~Serre, L.~Wolf, S.~Bileschi, M.~Riesenhuber, and T.~Poggio, ``Robust object
  recognition with cortex-like mechanisms,'' \emph{IEEE Trans. on Pattern
  Analysis and Machine Intelligence}, vol.~29, no.~3, pp. 411--26, Mar. 2007.

\bibitem{Boureau2010}
Y.-L. Boureau, F.~Bach, Y.~LeCun, and J.~Ponce, ``Learning mid-level features
  for recognition,'' in \emph{IEEE Computer Society Conference on Computer
  Vision and Pattern Recognition (CVPR)}, Jun. 2010, pp. 2559--2566.

\bibitem{lee2009unsupervised}
H.~Lee, Y.~Largman, P.~Pham, and A.~Y. Ng, ``Unsupervised feature learning for
  audio classification using convolutional deep belief networks,'' in
  \emph{Advances in Neural Information Processing System (NIPS)}, 2009.

\bibitem{Anden2013}
\BIBentryALTinterwordspacing
J.~And{\'e}n and S.~Mallat, ``Deep scattering spectrum,'' \emph{IEEE Trans.
  Signal Processing (submitted)}, 2014. [Online]. Available:
  \url{http://arxiv.org/abs/1304.6763}
\BIBentrySTDinterwordspacing

\bibitem{Abdel-Hamid2013}
O.~Abdel-Hamid, L.~Deng, and D.~Yu, ``Exploring convolutional neural network
  structures and optimization techniques for speech recognition,'' in
  \emph{Proc. INTERSPEECH 2013, 14th Annual Conf. of the ISCA}, Lyon, France,
  2013.

\bibitem{Hinton2012}
G.~Hinton, L.~Deng, D.~Yu, G.~Dahl, A.~Mohamed, N.~Jaitly, A.~Senior,
  V.~Vanhoucke, P.~Nguyen, T.~Sainath, and B.~Kingsbury, ``Deep neural networks
  for acoustic modeling in speech recognition: The shared views of four
  research groups,'' \emph{IEEE Signal Processing Magazine}, vol.~29, no.~6,
  pp. 82--97, 2012.

\bibitem{Smale2009}
S.~Smale, L.~Rosasco, J.~Bouvrie, A.~Caponnetto, and T.~Poggio, ``Mathematics
  of the neural response,'' \emph{Foundations of Computational Mathematics},
  vol.~10, no.~1, pp. 67--91, Jun. 2009.

\bibitem{Mallat2012}
S.~Mallat, ``Group invariant scattering,'' \emph{Communications on Pure and
  Applied Mathematics}, vol.~65, no.~10, pp. 1331--1398, 2012.

\bibitem{Poggio2003}
T.~Poggio and S.~Smale, ``The mathematics of learning: Dealing with data,''
  \emph{Notices of the AMS}, vol.~50, no.~50, pp. 537--544, 2003.

\bibitem{Liao2013}
Q.~Liao, J.~Z. Leibo, and T.~Poggio, ``Learning invariant representations and
  applications to face verification,'' in \emph{Advances in Neural Information
  Processing System (NIPS)}, 2013.

\bibitem{Schatz2013}
T.~Schatz, V.~Peddinti, F.~Bach, A.~Jansen, H.~Hermansky, and E.~Dupoux,
  ``Evaluating speech features with the minimal-pair {ABX} task: Analysis of
  the classical {MFC/PLP} pipeline,'' in \emph{Proc. INTERSPEECH 2013, 14th
  Annual Conf. of the ISCA}, Lyon, France, Aug. 2013.

\bibitem{O'Shaughnessy2013}
D.~O'Shaughnessy, ``Acoustic analysis for automatic speech recognition,''
  \emph{Proceedings of the IEEE}, vol. 101, no.~5, pp. 1038--1053, May 2013.

\bibitem{Stern2012}
R.~M. Stern and N.~Morgan, ``Hearing is believing: Biologically inspired
  methods for robust automatic speech recognition,'' \emph{IEEE Signal
  Processing Magazine}, vol.~29, no.~6, pp. 34--43, Nov. 2012.

\bibitem{Allen2009}
J.~B. Allen and F.~Li, ``{Speech perception and cochlear signal processing},''
  \emph{IEEE Signal Processing Magazine}, vol.~26, no.~4, pp. 73--77, Jul.
  2009.

\bibitem{Greenberg1997}
S.~Greenberg and B.~Kingsbury, ``The modulation spectrogram: in pursuit of an
  invariant representation of speech,'' in \emph{IEEE International Conference
  on Acoustics, Speech, and Signal Processing (ICASSP)}, vol.~3, 1997, pp.
  1647--1650.

\bibitem{HermanskySharma1998}
H.~Hermansky and S.~Sharma, ``{TRAPS}-classifiers of temporal patterns,'' in
  \emph{ISCA International Conference on Spoken Language Processing (ICSLP)},
  Sydney, Australia, Dec. 1998.

\bibitem{Chi2005}
T.~Chi, P.~Ru, and S.~A. Shamma, ``Multiresolution spectrotemporal analysis of
  complex sounds,'' \emph{The Journal of the Acoustical Society of America},
  vol. 118, no.~2, Aug. 2005.

\bibitem{Mesgarani2006}
N.~Mesgarani, M.~Slaney, and S.~Shamma, ``{Discrimination of speech from
  nonspeech based on multiscale spectro-temporal modulations},'' \emph{IEEE
  Transactions on Audio, Speech and Language Processing}, vol.~14, no.~3, pp.
  920--930, May 2006.

\bibitem{Kleinschmidt2003}
M.~Kleinschmidt, ``Localized spectro-temporal features for automatic speech
  recognition,'' in \emph{Proc. ISCA INTERSPEECH - Eurospeech}, 2003.

\bibitem{Bouvrie2008}
J.~Bouvrie, T.~Ezzat, and T.~Poggio, ``Localized spectro-temporal cepstral
  analysis of speech,'' in \emph{Proc. IEEE International Conference on
  Acoustics, Speech and Signal Processing (ICASSP)}.\hskip 1em plus 0.5em minus
  0.4em\relax IEEE, Mar. 2008, pp. 4733--4736.

\bibitem{Heckmann2011}
M.~Heckmann, X.~Domont, F.~Joublin, and C.~Goerick, ``A hierarchical framework
  for spectro-temporal feature extraction,'' \emph{Speech Communication},
  vol.~53, no.~5, pp. 736--752, May 2011.

\bibitem{Abdel-Hamid2012}
O.~Abdel-Hamid, A.-R. Mohamed, H.~Jiang, and G.~Penn, ``Applying convolutional
  neural networks concepts to hybrid {NN-HMM} model for speech recognition,''
  in \emph{Proc. IEEE International Conference on Acoustics, Speech and Signal
  Processing (ICASSP)}, Mar. 2012, pp. 4277--4280.

\bibitem{Sainath2013}
T.~N. Sainath, A.~Mohamed, B.~Kingsbury, and B.~Ramabhadran, ``Deep
  convolutional neural networks for {LVCSR},'' in \emph{Proc. IEEE
  International Conference on Acoustics, Speech and Signal Processing
  (ICASSP)}, May 2013, pp. 8614--8618.

\bibitem{Yu2013a}
D.~Yu, M.~L. Seltzer, J.~Li, J.-T. Huang, and F.~Seide, ``Feature learning in
  deep neural networks - studies on speech recognition tasks,'' in \emph{Proc.
  International Conference on Learning Representations (ICLR)}, 2013.

\bibitem{Roy2009a}
B.~C. Roy, M.~C. Frank, and D.~Roy, ``Exploring word learning in a high-density
  longitudinal corpus,'' in \emph{The Annual Meeting of the Cognitive Science
  Society (CogSci '09)}, 2009, pp. 2106--2111.

\bibitem{Niyogi1998}
P.~Niyogi, F.~Girosi, and T.~Poggio, ``Incorporating prior information in
  machine learning by creating virtual examples,'' \emph{Proceedings of the
  IEEE}, vol.~86, no.~11, pp. 2196--2209, 1998.

\bibitem{Jaitly2013}
N.~Jaitly and G.~Hinton, ``Vocal tract length perturbation {(VTLP)} improves
  speech recognition,'' \emph{Proc. ICML Workshop on Deep Learning for Audio,
  Speech and Language Processing}, 2013.

\bibitem{Maaten2008}
L.~V. der Maaten and G.~Hinton, ``Visualizing data using {t-SNE},''
  \emph{Journal of Machine Learning Research}, vol.~9, no.~11, 2008.

\bibitem{Halberstadt1997}
A.~Halberstadt and J.~Glass, ``Heterogeneous acoustic measurements for phonetic
  classification,'' in \emph{EUROSPEECH - 5th European Conference on Speech
  Communication and Technology}, Rhodes, Greece, Sep. 1997.

\bibitem{Tacchetti2013}
A.~Tacchetti, P.~K. Mallapragada, M.~Santoro, and L.~Rosasco, ``{GURLS}: A
  least squares library for supervised learning,'' \emph{Journal of Machine
  Learning Research}, vol.~14, Oct. 2013.

\bibitem{Karsmakers2007}
P.~Karsmakers, K.~Pelckmans, J.~A. Suykens, and {Hugo Van hamme}, ``Fixed-size
  kernel logistic regression for phoneme classification,'' in \emph{Proc.
  INTERSPEECH 2007, 8th Annual Conf. of the ISCA}, Antwerp, Belgium, Aug. 2007,
  pp. 78--81.

\bibitem{Rifkin2007}
R.~Rifkin, K.~Schutte, M.~Saad, J.~Bouvrie, and J.~R. Glass, ``Noise robust
  phonetic classification with linear regularized least squares and
  second-order features,'' in \emph{Proc. IEEE International Conference on
  Acoustics, Speech and Signal Processing (ICASSP)}, Honolulu, Hawaii, Apr.
  15-20 2007.

\bibitem{HTK09}
(2009) {HTK Speech Recognition Toolkit, HTK version 3.4.1}.
  \url{http://htk.eng.cam.ac.uk/}.

\bibitem{Chang2009}
H.-A. Chang and J.~R. Glass, ``Hierarchical large-margin {Gaussian} mixture
  models for phonetic classification,'' in \emph{IEEE Workshop on Automatic
  Speech Recognition \& Understanding (ASRU)}, 2007.

\end{thebibliography}
\end{document}